# Information Divergence and Endogenous Evolution of Asymmetric Information

**Ratna K. Shrestha[1]*, Gaurav Subedi[2], and Ayush Shrestha[3]**

[1]Sauder School of Business, University of British Columbia, Vancouver, BC, Canada

[2]RALP Technologies, Vancouver, BC, Canada

[3]Global Relay Communications, Vancouver, BC, Canada

**Abstract**

Standard theories trace information asymmetry to privately held information, heterogeneous signals, or costly information acquisition. We identify a distinct mechanism: otherwise identical rational agents observing a common Markov state at asynchronous times can hold different predictive distributions because their information differs in age. When the transition kernel separates information ages, independent Bernoulli observation arrivals make such divergence recur infinitely often almost surely, even with identical arrival probabilities. When agents instead choose costly updating intensities, independent update realizations can regenerate heterogeneous information ages among ex ante identical agents. In a stationary Gaussian AR(1) environment, the consequences depend on common staleness as well as the age gap: holding the gap fixed, older information raises individual uncertainty while reducing expected squared forecast disagreement and the value of certifying relative freshness. This generates stale consensus: low disagreement despite high uncertainty. With multiple ordered ages, costly certification supports partial-unraveling equilibria in which fresher types certify and sufficiently stale types pool. Under geometric information ages, more frequent updating weakly lowers the smallest and largest equilibrium disclosure cutoffs wherever positive-certification equilibria exist. The framework has implications for credit, insurance, financial markets, and multi-agent AI.

**Keywords**: Information Age, Asymmetric Information, Asynchronous Observation, Information Divergence, Stale Consensus, Disclosure, Certification.

JEL: D82, D83, D84, G14

---

*Corresponding Author: ratna.shrestha@ubc.ca or shrestha.rk11@gmail.com

## 1. Introduction

Information asymmetry underpins theories of adverse selection, moral hazard, contract design, and financial market efficiency. Classical models attribute asymmetric information to privately held information, including hidden characteristics and hidden actions (Akerlof, 1970; Spence, 1973; Rothschild and Stiglitz, 1976; Holmström, 1979). Subsequent work studies endogenous information acquisition (Grossman and Stiglitz, 1980; Veldkamp, 2006; Zhong, 2022), limited information-processing capacity under rational inattention (Sims, 2003; Steiner, Stewart, and Matějka, 2017), and the emergence of asymmetric information through acquisition or investment decisions (Gonzalez, 2004; Barron and Ni, 2008).

This paper identifies a distinct mechanism generating informational differences. In the baseline model, agents do not choose to acquire information. Instead, they receive fully revealing observations of a common payoff-relevant Markov state at potentially different times. The mechanism operates through *information age*, the time elapsed since an agent's most recent observation. Together, the last observed state and its age determine the agent's predictive distribution over the current state. When the transition kernel is age-separating—so observations of different ages generate distinct predictive distributions—asynchronous observation produces what we call *information divergence*. Asymmetric information can thus emerge among rational agents with identical initial information, forecasting models, and observation technologies. Under independent, nondegenerate Bernoulli arrivals, this divergence recurs infinitely often almost surely, even when all agents face the same arrival probability.

Next, we endogenize information updating by allowing agents to choose the intensity of a stochastic observation process. When updating is costly, agents trade off the resource cost of maintaining a higher refresh rate against the expected loss from acting on stale information. Unlike information-acquisition models in which agents choose the precision or informativeness of a signal (e.g., Cai, 2019; Zhong, 2022), an update in our framework fully reveals the contemporaneous state and resets information age to zero. The relevant choice is therefore the *intensity* of updating, rather than the precision of information acquired. Thus asynchronous information states can arise not only from exogenous information arrival but also from optimal information updating behavior.

Under state-independent observation timing, predictive distributions depend on the last observed state and its age, regardless of whether updating intensity is exogenous or chosen. Specializing to a mean-reverting Gaussian AR(1) process, we show that the consequences depend not only on the age gap between agents but also on the common staleness of their observations. Holding the age gap fixed, greater common staleness raises each agent's predictive uncertainty but reduces the disagreement between their predictive forecasts. As common staleness grows, disagreement converges to zero while individual uncertainty approaches its stationary variance. This *stale-consensus effect* explains why agreement need not indicate better information: agents may issue similar forecasts because both have current information or because both rely on sufficiently stale observations. The model therefore provides a structural reason why forecast disagreement and uncertainty can move in opposite directions.

Common staleness also shapes incentives to disclose information age. When a counterparty values forecast accuracy, the value of certifying relative freshness declines as both observations become stale. In a binary fresh-versus-stale setting, sufficiently large common staleness eliminates the fresher type's incentive to separate. With multiple ordered information ages, costly certification yields a threshold equilibrium in which sufficiently fresh agents certify their exact age while sufficiently stale agents pool. Any positive certification cost precludes full disclosure when information ages have unbounded support. Under geometric information ages, we derive the conditions for threshold equilibria and full pooling, show that they can coexist, and characterize how the equilibrium cutoffs change with updating frequency. Stale consensus and limited disclosure can therefore arise together: commonly old information can produce both low forecast disagreement and weak incentives to establish freshness. Figure 1 summarizes the mechanism.

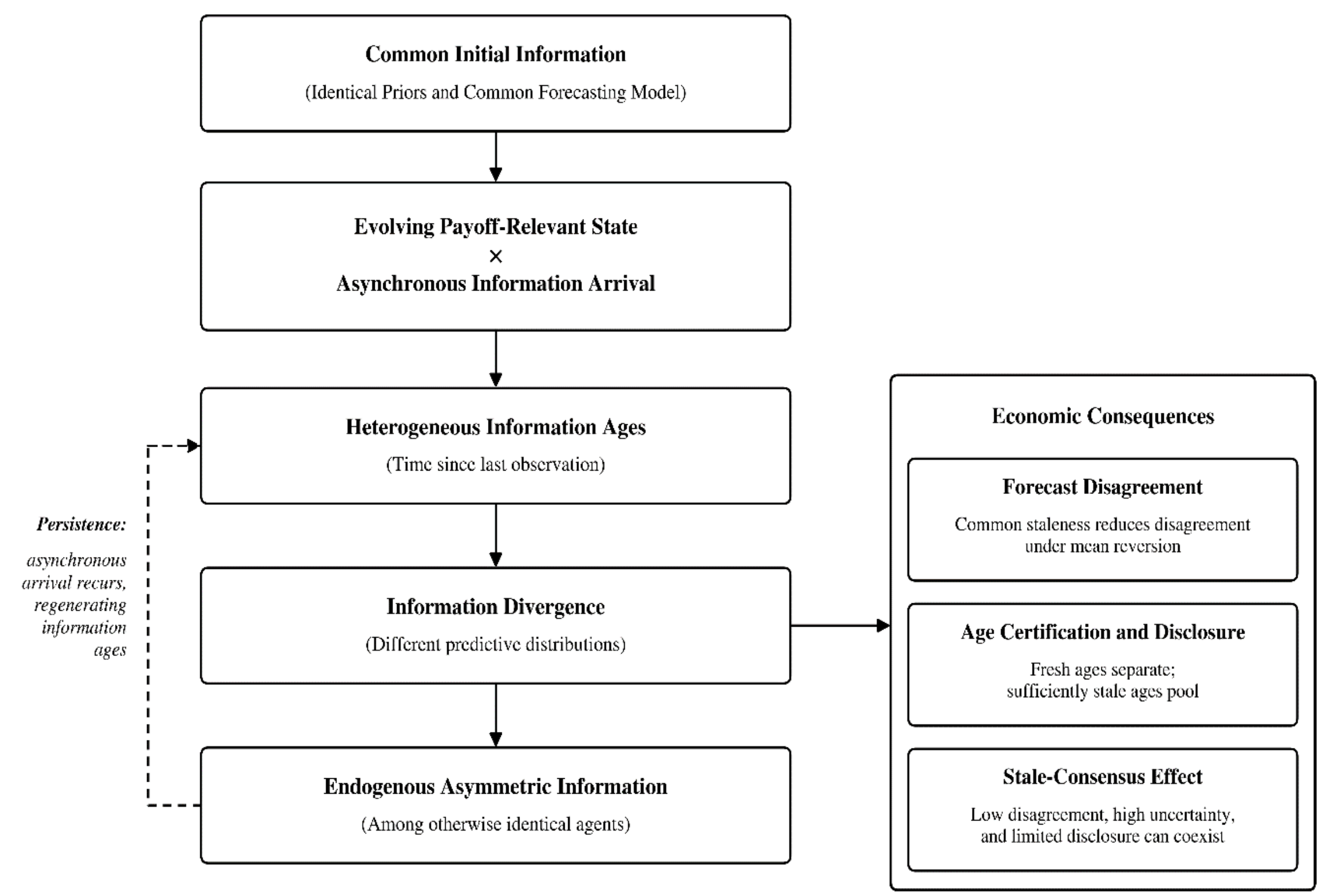

**Figure 1. Mechanism for the Endogenous Emergence and Economic Consequences of Asymmetric Information**

The paper makes four main contributions. First, it identifies asynchronous observation of an evolving common state as a source of asymmetric information among ex ante identical agents. Second, it establishes almost-sure recurrence under independent Bernoulli arrivals and extends this result to optimally chosen interior updating intensities. Third, it shows how common staleness can increase predictive uncertainty while reducing expected forecast disagreement and the value of relative freshness. Fourth, it characterizes information-age certification equilibria in which state dynamics determine valuation differences across types. These results connect information timing to the emergence, observable consequences, and disclosure of asymmetric information.

Section 2 reviews the related literature. Section 3 presents the model and its information structure. Section 4 establishes the emergence and recurrence of information divergence and endogenizes updating. Section 5 develops the linear-Gaussian specialization and examines common staleness, forecast disagreement, and information-age certification. Section 6 discusses applications to credit, insurance, financial markets, and multi-agent AI. Section 7 concludes.

## 2. Related Literature

This paper relates to several strands of research in information economics, communication engineering, and information systems. Together, these literatures examine how information is acquired, processed, transmitted, and updated; how its timing and freshness affect beliefs and decisions; and how informational differences shape economic and strategic outcomes.

### *2.1 Classical Information Economics*

Classical information economics traces asymmetric information to privately held characteristics or actions: hidden types, hidden actions, and costly signals. Akerlof (1970), Spence (1973), Rothschild and Stiglitz (1976) build foundational models of adverse selection, signaling, and screening on such private information. A parallel literature on principal-agent problems examines hidden actions under imperfect monitoring (Holmström, 1979; Grossman and Hart, 1983), while subsequent work studies how prices, advertising, and other costly actions can serve as signals that communicate privately held information (Milgrom and Roberts, 1986). Across these models, informational differences originate in private types, actions, or signals that are exogenously given.

### *2.2 Information Acquisition, Rational Inattention, and Endogenous Asymmetry*

A second strand traces informational differences to choices and constraints in acquiring and processing information. Grossman and Stiglitz (1980) show that costly information acquisition limits the extent to which prices can fully reveal underlying information. Subsequent work studies strategic acquisition, including coordination motives and complementarities in learning (Hellwig and Veldkamp, 2009; Myatt and Wallace, 2012) and increasing returns in information production that generate common information across agents (Veldkamp, 2006). The rational-inattention literature instead models costly limits on information processing and endogenous choice of information structure (Sims, 2003; Matějka and McKay, 2015; Steiner, Stewart, and Matějka, 2017), including in strategic environments with endogenous information acquisition (Hébert and La'O, 2023).

Several studies show that such choices can make asymmetric information itself endogenous. Gonzalez (2004) shows that an agent's unobservable randomized investment strategy can generate private information in a screening environment, and Barron and Ni (2008) show that portfolio size shapes incentives to acquire costly information about foreign assets. In dynamic settings, Cai (2019) studies how time-varying fundamental uncertainty affects acquisition incentives, and Zhong (2022) characterizes optimal acquisition when information accumulates over time. In these models, informational differences arise from what agents choose to learn, or how precisely they learn it. In ours, they arise even without any choice, because every observation is exact and differs only in timing. When information updating is chosen (Section 4.3), the choice concerns updating intensity rather than precision.

Sticky-information models study a closely related friction: infrequent refreshing of information (Mankiw and Reis, 2002; Reis, 2006). Building on the staggered-adjustment logic of Calvo (1983), these models shift the friction from price adjustment to information, and survey forecasts provide empirical support (Coibion and Gorodnichenko, 2012, 2015). Because agents update at different dates, sticky information also generates cross-sectional dispersion in expectations. Mankiw, Reis, and Wolfers (2003) use this mechanism to explain disagreement about inflation expectations, and Andrade and Le Bihan (2013) document that professional forecasters update infrequently and disagree persistently. We share the premise that differences in information vintage produce different forecasts, but depart from this literature in three ways. First, rather than using dispersion as an input to aggregate dynamics, we treat it as asymmetric information and characterize when it arises and recurs among ex ante identical agents. Second, we show that its observable consequences depend on common staleness as well as the age gap. Third, we treat information age as a private type and study incentives to certify it. On the updating side, the Bernoulli arrival process in Theorem 2 has a constant per-period hazard, as in Calvo (1983), and Section 4.3 endogenizes its intensity in the spirit of Reis (2006). Unlike Reis's inattentive producers, who update on a deterministic schedule, our agents choose the intensity of a stochastic updating technology; this randomness is what desynchronizes ex ante identical agents.

### *2.3 Dynamic Information, Learning, and Information Design*

A third strand studies how information evolves over time through learning, disclosure, and strategic interaction. Bayesian persuasion and dynamic information design examine how a strategically informed sender determines the timing and content of information revealed to a receiver in order to shape the receiver's beliefs and actions (Kamenica and Gentzkow, 2011; Renault, Solan, and Vieille, 2017; Ely, 2017; Smolin, 2021; Ball, 2023). Daley and Green (2012) study a market in which a seller's private information about asset quality is gradually corroborated by public news arriving over time; the asymmetry originates in the seller's initial private type, with news arrival governing how quickly it resolves. A related literature on dynamic games characterizes equilibrium beliefs when agents' private information itself evolves over time, rather than being designed by a strategic sender (Fershtman and Pakes, 2012; Ouyang, Tavafoghi, and Teneketzis, 2025).

### *2.4 Disclosure, Certification, and Unraveling*

A related literature studies voluntary disclosure when privately informed agents can credibly reveal information. When disclosure is costless and the sender is known to be informed, skepticism about silence produces full unraveling, with only the worst type left undisclosed (Grossman, 1981; Milgrom, 1981). Unraveling is incomplete when disclosure costs induce types below a threshold to withhold information (Jovanovic, 1982; Verrecchia, 1983), or when uncertainty about whether the sender is informed makes silence indistinguishable from ignorance (Dye, 1985).

Our disclosure results build on this logic but change the nature of the private type: the age of an otherwise exact observation of an evolving state. Valuation differences between types arise from state dynamics: under mean reversion, the value gap between adjacent ages shrinks as ages grow. Any positive certification cost therefore precludes full disclosure. In threshold equilibria, fresher ages certify while sufficiently stale ages pool (Section 5.4). Equilibrium cutoffs depend on certification cost, state persistence, and, under geometric ages, observation frequency.

### *2.5 Age of Information*

The closest conceptual connection to our framework outside economics is the engineering literature on Age of Information (AoI). Kaul, Yates, and Gruteser (2012) introduce information age as a measure of information freshness, defined as the time elapsed since the generation of the most recently received update. Subsequent research examines how communication networks, queueing systems, scheduling algorithms, and transmission policies affect information freshness (Yates et al., 2021). Related ideas appear in distributed state estimation, where multiple sensors estimate a common process from asynchronous, delayed observations (e.g., Mitra et al., 2022).

In the AoI literature, information age primarily measures the freshness of information in a communication system and itself is an object of optimization. Here, information age is an economic informational state variable that determines how an agent's last observation maps into a predictive distribution over the current payoff-relevant state (see also Subedi and Shrestha, 2026).

### *2.6 Forecast Disagreement and Uncertainty*

A separate empirical literature asks whether disagreement among forecasters is a reliable measure of uncertainty. Zarnowitz and Lambros (1987) distinguish consensus forecasts from individual forecasters' uncertainty. Bomberger (1996) examines disagreement as an empirical measure of uncertainty, while Lahiri and Sheng (2010) derive conditions under which the two concepts need not coincide. Our framework provides a structural explanation for that gap. Holding the information-age gap fixed, increasing common staleness causes forecast disagreement to decline even as each agent's predictive uncertainty rises toward the stationary variance, because stale conditional means converge toward the same long-run mean. This stale-consensus effect implies that disagreement can understate uncertainty precisely when the underlying information is old.

## 3. The Model

To develop the mechanism through which asymmetric information emerges endogenously, the model combines two essential ingredients: (i) an evolving payoff-relevant state and (ii) asynchronous information observation across otherwise identical agents.

### *3.1 Economic Environment*

Consider a discrete-time economy $t = 0, 1, 2, \ldots$, with a finite number of rational agents $i = 1, 2, \ldots, N$, (where $N \geq 2$). A payoff-relevant state variable $\{X_t\}_{t\geq 0}$, taking values in a measurable state space $(\mathcal{X}, \mathcal{B})$, evolves according to a time-homogeneous Markov process with transition kernel $Q$: $X_{t+1} \sim Q(\cdot \mid X_t)$, where $Q(B \mid x) = \Pr(X_{t+1} \in B \mid X_t = x), B \in \mathcal{B}, and\ x \in \mathcal{X}.$

The state may represent a borrower's creditworthiness, a firm's fundamentals, an individual's health status, a worker's productivity, or any other payoff-relevant economic characteristic that influences agents' decisions and payoffs. At $t = 0$, every agent observes the initial state $X_0$, shares common knowledge of $Q$, and uses the same forecasting rule. The Markov specification includes, among others, random walks, autoregressive processes, and finite-state Markov chains.

### *3.2 Information Structure*

Following the initial period, agents observe the true state only intermittently. Let $\mathcal{S}_i(t) \subseteq \{0, 1, 2, \ldots, t\}$, denote the set of dates at which agent $i$ has observed the state by date $t$, with $0 \in \mathcal{S}_i(t)$. Agent $i$'s full observation history at date $t$ is $I_i(t) = \{(s, X_s) : s \in \mathcal{S}_i(t)\}$, together with common knowledge of the transition kernel $Q$. Thus $I_i(t)$ contains every state observation previously received by agent $i$, not merely the most recent one. Define $\tau_i(t) = max\ \mathcal{S}_i(t)$ as the date of agent $i$'s most recent observation of the true state.

Agent $i$'s realized observation history, including both the dates and values of its observations, is privately observed by that agent. Agents know the common transition kernel and the observation-arrival process, but do not observe other agents' realized observation histories, nor the recency of their observations. Sections 5.3–5.4 allow the recency of information to be verifiably disclosed.

**Assumption 1 (Exogenous Observation Timing).** Observation times are independent of the payoff-relevant state process $X_t$. Hence, the occurrence or absence of an observation conveys no information about the state beyond the value revealed when an observation occurs.

Informational heterogeneity therefore arises from differences in observation timing rather than from heterogeneous priors, signals, or forecasting technologies. Because $X_t$ follows a time-homogeneous Markov process, the predictive content of an agent's observation history is summarized by the most recently observed state and the time elapsed since that observation.

*3.3 Information Age and Effective Information State*

**Definition 1 (Information Age)**

Following Kaul, Yates, and Gruteser (2012), the information age of agent *i* at time *t* is $A_i(t) = t - \tau_i(t)$. Thus, $A_i(t) = 0$ indicates that the agent observes the current state, while $A_i(t) > 0$ measures the elapsed time since the agent's most recent observation. Information age evolves according to:

$$A_i(t+1) = \begin{cases} 0, & \text{if agent } i \text{ receives fresh information at } t+1, \\ A_i(t)+1, & \text{otherwise.} \end{cases}$$

**Definition 2 (Asynchronous Information Observation)**

Information observation is asynchronous at date *t* if there exists $i \neq j$ such that $\tau_i(t) \neq \tau_j(t)$. Equivalently, $A_i(t) \neq A_j(t)$. Information age alone does not completely characterize an agent's beliefs, however. The predictive distribution depends jointly on the *age* of the information and the *value* of the most recently observed state. The relevant effective information state is therefore the pair $(X_{\tau_i(t)}, A_i(t))$.

Because agents observe a common state process, two agents with the same information age at a given date necessarily have the same most recent observation date and the same last observed state. Hence cross-agent differences in effective information states arise if and only if their information ages differ.

**Proposition 1 (Information Set Representation)**

Under Assumption 1 and the time-homogeneous Markov transition kernel $Q$, agent $i$'s full observation history, $I_i(t)$, contains no more predictive information about $X_t$ than the pair $(X_{\tau_i(t)}, A_i(t))$—the most recently observed state and its age. Formally, for every measurable set $B$, $\Pr(X_t \in B \mid I_i(t)) = Q^{A_i(t)}(B \mid X_{\tau_i(t)})$. Hence, $(X_{\tau_i(t)}, A_i(t))$ is a sufficient statistic for agent $i$'s predictive distribution of $X_t$.

**Proof.** By definition, $\tau_i(t)$ is the date of agent $i$'s most recent observation. Conditional on $X_{\tau_i(t)}$, the Markov property implies that earlier observations $\{X_s : s \in \mathcal{S}_i(t), s < \tau_i(t)\}$ contain no additional information about $X_t$. Time homogeneity then implies that the distribution of $X_t$ depends on the elapsed time since the most recent observation only through $A_i(t)$. Therefore, $\Pr(X_t \in B \mid I_i(t)) = Q^{A_i(t)}(B \mid X_{\tau_i(t)})$. □

*3.4 Optimal Forecasting under Asynchronous Information*

For a real-valued state with finite second moment and squared-error loss, agent $i$'s optimal forecast given $I_i(t)$ is the conditional mean: $\hat{X}_i(t) = \mathbb{E}[X_t \mid I_i(t)]$. By Proposition 1, the full observation history is summarized for forecasting purposes by the most recently observed state and its age. Hence, $\hat{X}_i(t) = \mathbb{E}[X_t \mid X_{\tau_i(t)}, A_i(t)]$.

Forecasts can nonetheless differ when agents condition on observations of different vintages, so forecast heterogeneity reflects differences in available information rather than in priors or forecasting rules.

*3.5 Information Divergence*

**Definition 3 (Information Divergence)**

Information divergence occurs at date $t$ if there exist agents $i \neq j$ such that their predictive distributions over the current state differ: $\mu_i(t) \equiv \Pr\big(X_t \in \cdot \,\big|\, I_i(t)\big) \neq \Pr\Big(X_t \in \cdot \,\Big|\, I_j(t)\Big) \equiv \mu_j(t)$. Information divergence concerns predictive distributions rather than information ages or effective information states alone. Different information ages need not, in general, imply different predictive distributions; Definition 4 states the condition under which they do.

**Definition 4 (Age-Separating Transition Kernel)**

The transition kernel $Q$ is age-separating if for any $a \neq a'$ and $x, x' \in \mathcal{X}$, $Q^{a}(\cdot| \, x) \neq Q^{a'}(\cdot| \, x')$. Thus observations of different ages induce distinct predictive distributions, regardless of the realized values of the most recent observations.

## 4. Theoretical Results

### *4.1 Dynamic Emergence of Asymmetric Information*

**Theorem 1 (Emergence of Asymmetric Information)**

Suppose agents begin with common information at $t = 0$, share the same optimal forecasting rule, and the transition kernel $Q$ is age-separating. If at some date $t \geq 1$, $A_i(t) \neq A_j(t)$ for some agents $i \neq j$, then $\mu_i(t) \neq \mu_j(t)$.

**Proof.** By Proposition 1, $\mu_i(t) = Q^{A_i(t)}\left(\cdot \mid X_{\tau_i(t)}\right)$ and similarly for agent $j$. Since $A_i(t) \neq A_j(t)$ for some $i \neq j$ and $Q$ is age-separating, the two predictive distributions are distinct. By Definition 3, information divergence occurs at $t$. □

Theorem 1 isolates our central mechanism: asynchronous observation of an evolving payoff-relevant state can generate asymmetric information among otherwise identical rational agents.

### *4.2 Persistence of Information Asymmetry*

We now ask whether information divergence is merely transitory or can recur indefinitely under stochastic information arrival.

**Theorem 2 (Almost-Sure Persistence under Stochastic Information Arrival)**

For each agent $i$, let

$$Z_i(t) \sim i.i.d. \, \text{Bernoulli}(p_i), \; p_i \in (0, 1),$$

where $Z_i(t) = 1$ indicates that agent $i$ receives fresh information at date $t$, and $p_i$ is the exogenously given information arrival probability. Suppose observation arrivals are independent across agents and over time, and the transition kernel is age-separating. Then heterogeneous information ages occur infinitely often almost surely. Consequently, information divergence—and hence asymmetric information—recurs infinitely often almost surely.

**Proof.** Fix any pair $i \neq j$ and define $E_t^{ij} = \{Z_i(t) \neq Z_j(t)\}$, the event that exactly one of the two agents receives a fresh observation at date *t*. Whenever $E_t^{ij}$ occurs, one agent receives fresh information and the other does not. Thus, without loss of generality, $Z_i(t) = 1 \Longrightarrow A_i(t) = 0$, while $Z_j(t) = 0 \Longrightarrow A_j(t) \geq 1$. Hence, $E_t^{ij} \Longrightarrow A_i(t) \neq A_j(t)$.

Since $p_i, p_j \in (0,1)$, each event has probability $\Pr(E_t^{ij}) = p_i(1 - p_j) + p_j(1 - p_i) > 0$, so that $\sum_{t=1}^{\infty} \Pr(E_t^{ij}) = \infty$. Moreover, because information arrivals are independent over time, the events $\{E_t^{ij}\}_{t\geq 1}$ are mutually independent. The second Borel–Cantelli lemma therefore implies $\Pr(E_t^{ij}$ infinitely often$) = 1$. Thus, $A_i(t) \neq A_j(t)$ occurs infinitely often almost surely. Because the transition kernel is age-separating, Theorem 1 implies that information divergence occurs at each such date. Hence asymmetric information recurs infinitely often almost surely. □

Theorem 2 requires no ex ante heterogeneity across agents. If $p_i = p_j = p \in (0, 1)$, then $\Pr(E_t^{ij}) = 2p(1-p) > 0$. Thus even ex ante identical agents, facing identical information-arrival processes and sharing the same forecasting model, become informationally differentiated infinitely often almost surely, solely because their realized observation times differ.

### *4.3 Endogenous Information Updating*

The preceding analysis treats information arrival as exogenous. We now allow agents to choose how intensively to update. Updating is costly but reduces the expected loss from acting on stale information. This trade-off parallels that in Reis (2006), where producers optimally choose how often to acquire and process information. Unlike in Reis (2006), where inattentive producers update on a deterministic schedule, we model updating as a costly monitoring effort whose success is stochastic: the agent chooses the per-period probability that a fresh observation is obtained, but not the exact timing of updates. This randomness is what allows ex ante identical agents to desynchronize; under deterministic schedules, identical agents starting from common information would update simultaneously.

Agent $i$ chooses an updating intensity $u_i \in (0,1]$.[1] Conditional on this intensity, a fresh observation becomes available in each period with probability $u_i$, independently across periods and agents. Thus $u_i$ determines the expected frequency with which agent *i*'s information is refreshed while the realized timing of individual updates remains stochastic. Let

[1] The boundary $u_i = 1$ is included, and corresponds to updating every period; $u_i = 0$, never updating, is excluded since it sends expected information age to infinity.

$$Z_i(t) \sim i.i.d.\,\text{Bernoulli}(u_i), \qquad u_i \in (0,1],$$

independently across agents and over time, where $Z_i(t) = 1$ denotes a fresh observation of $X_t$ at date *t*. A higher value of $u_i$ therefore corresponds to more frequent updating.[2] The resulting information-age process satisfies

$$A_i(t+1) = \begin{cases} 0, & Z_i(t+1) = 1, \\ A_i(t) + 1, & Z_i(t+1) = 0. \end{cases}$$

For any $u_i \in (0,1]$, the information-age Markov chain has stationary distribution $Pr(A_i = a \mid u_i) = u_i(1 - u_i)^a$, $a = 0, 1, 2, 3 \ldots$ Information has age *a* if a fresh observation occurred *a* periods ago and no update thereafter. Consequently, $\mathbb{E}[A_i \mid u_i] = \sum_{a=0}^{\infty} a\, u_i(1 - u_i)^a = (1 - u_i)/u_i$, which is strictly decreasing in the updating intensity.

For each information age $a$, define $\phi_i(a) = \mathbb{E}\left[\ell_i\big(X_t, \hat{X}_i(t)\big) \mid A_i(t) = a\right]$, where $\ell_i(\cdot,\cdot)$ is the pointwise loss from imperfect forecasting of the current state. Thus $\phi_i(a)$ is agent *i*'s expected forecasting loss conditional on its information being *a* periods old. Assume that the expected loss from stale information is nondecreasing in information age, $\phi_i(a+1) \geq \phi_i(a)\ \forall a \geq 0$. Under updating intensity $u_i$, agent *i*'s long-run expected forecasting loss is $L_i(u_i) = \sum_{a=0}^{\infty} u_i(1 - u_i)^a \phi_i(a)$.

**Lemma 1.** Suppose the stationary loss schedule $\phi_i(a)$ is independent of $u_i$, nondecreasing and bounded in information age *a*. Then, $L_i(u_i)$ is nonincreasing and convex in $u_i$ on (0, 1].

**Proof.** Under Bernoulli updating with intensity $u_i \in (0,1]$, the stationary distribution of information age is $Pr(A_i = a \mid u_i) = u_i(1 - u_i)^a$, $a = 0, 1, 2, 3 \ldots.$ with cumulative distribution function (CDF) $F_i(a;\, u_i) = \Pr(A_i \leq a \mid u_i) = 1 - (1 - u_i)^{a+1}$.

For each $a \geq 0$, $F_i(a;\, u_i)$ is nondecreasing in $u_i$. Thus a higher $u_i$ shifts the stationary distribution of information age towards lower ages in the first order stochastic dominance sense. Let $\Delta\phi_i(a) = \phi_i(a+1) - \phi_i(a)$. Using the tail-sum representation and $1 - F_i(a;\, u_i) = (1 - u_i)^{a+1}$, we get $L_i(u_i) = \phi_i(0) + \sum_{a=0}^{\infty} \Delta\phi_i(a)(1 - u_i)^{a+1}$.

[2] Calvo (1983) models adjustment opportunities using a continuous time Poisson process; we use its discrete-time analogue: Bernoulli arrivals. The analogy holds only for the stochastic and asynchronous nature of adjustment. Calvo's hazard is exogenous and age-independent; here, updating intensity is endogenous, while information age affects the agent's predictive distribution $\mu_i(t)$, and thus the value of updating.

Because $\Delta\phi_i(a) \geq 0$, $(1 - u_i)^{a+1}$ is non-increasing in $u_i$, and $\sum_{a=0}^{\infty} \Delta\, \phi_i(a) = \lim_{a\to\infty}\phi_i(a) - \phi_i(0)$ is finite, $L'_i(u_i) \leq 0$ and $L_i''(u_i) = \sum_{a=1}^{\infty} a\,(a+1)\, \Delta\phi_i(a)\,(1 - u_i)^{a-1} \geq 0$. □

Let $C_i(u_i)$ denote the direct resource cost of updating, and let $\lambda_i > 0$ measure the economic importance of forecast accuracy. Agent $i$ chooses $u_i$ to minimize $J_i(u_i) = \lambda_i L_i(u_i) + C_i(u_i)$, so that $u_i^* \in arg \min_{u_i \in (0,1]} J_i(u_i)$.

**Proposition 2 (Optimal Endogenous Information Updating)**

Suppose $\phi_i$ is nondecreasing and bounded, and $C_i$ is continuously differentiable and strictly convex on (0, 1]. Then $J_i$ is strictly convex on (0, 1], and an interior optimal updating intensity $u_i^*$, if it exists, is unique and satisfies $C'_i(u_i^*) = -\lambda_i L'_i(u_i^*)$. At the optimum, the marginal resource cost of additional updating intensity equals the marginal reduction in expected forecast loss from fresher information. If, in addition, $\lim_{u_i \downarrow 0} J_i'(u_i) < 0$ and $J_i'(1) > 0$, then a unique interior optimum $u_i^* \in (0,1)$ exists.

**Proof.** By Lemma 1, $L_i$ is convex, so $J_i = \lambda_i L_i + C_i$ is strictly convex because $C_i$ is. An interior minimizer therefore satisfies the first-order condition (FOC): $C_i'(u_i^*) = -\lambda_i L_i'(u_i^*)$. By Lemma 1, the right-hand side is nonnegative and represents the marginal benefit of additional updating intensity through reduced expected forecast loss. Strict convexity makes any point satisfying the FOC the unique global minimizer of $J_i$ on (0, 1].

If $\lim_{u_i \downarrow 0} J_i'(u_i) < 0$ and $J_i'(1) > 0$, continuity of $J_i'(u_i)$ implies that it crosses zero at some $u_i^* \in (0,1)$. Strict convexity makes this crossing unique. □

Proposition 2 implies that perfectly current information need not be optimal. When updating is sufficiently costly, the optimal intensity satisfies $u_i^* < 1$, so that the agent rationally accepts a positive probability of operating with stale information. Differences in $C_i(u_i)$, $\lambda_i$, or $\phi_i(a)$ can generate heterogeneous optimal updating intensities across agents.

Importantly, however, heterogeneity in $u_i^*$ is not required for information divergence. Ex ante identical agents facing identical costs and losses choose the same interior $u^*$. Because the updating technology is stochastic, independent realizations of their updating processes nonetheless generate different information ages.

**Corollary 1 (Endogenous Updating and Information Divergence)**

Suppose identical agents choose the same interior updating intensity $u^* \in (0,1)$, with update realizations independent across agents and over time. If the transition kernel is age-separating, then heterogeneous information ages, and hence information divergence, recur infinitely often almost surely.

**Proof.** The updating processes satisfy the conditions of Theorem 2 with $p_i = p_j = u^*$. Therefore, $\Pr\left(E_t^{ij}\right) = 2u^*(1 - u^*) > 0$. So different information ages occur infinitely often almost surely. By Theorem 1, each such age difference generates different predictive distributions. □

**5. Linear Gaussian Specialization and Economic Consequences**

The preceding section developed the framework under a general Markov process. This section specializes the analysis to a linear Gaussian AR(1) process and its limiting case, random walk.

*5.1 Predictive Uncertainty*

Suppose the payoff-relevant state follows the stationary AR(1) process: $X_{t+1} = \bar{x} + \rho(X_t - \bar{x}) + \varepsilon_{t+1}$, where $\bar{x}$ denotes the long-run mean, $\rho \in (-1, 1)$ is the autoregressive coefficient, and $\varepsilon_{t+1} \sim \text{i.i.d.}\, N(0, \sigma^2)$. Economically, $\sigma^2$ captures the intrinsic volatility of the underlying economic environment.

When $|\rho| < 1$, the state is stationary and mean-reverting around the long-run mean $\bar{x}$. A larger $|\rho|$ implies greater persistence, or slower decay towards $\bar{x}$, while a smaller $|\rho|$ implies faster decay. When $\rho > 0$, convergence towards $\bar{x}$ is monotonic in the forecast horizon, whereas when $\rho < 0$, the conditional mean oscillates around $\bar{x}$ with diminishing amplitude. At the unit-root boundary $\rho = 1$, the specification reduces to the random walk $X_{t+1} = X_t + \varepsilon_{t+1}$ for which stationarity no longer holds.

Iterating the state equation from agent *i*'s most recent observation yields $X_t = \bar{x} + \rho^{A_i(t)}[X_{\tau_i(t)} - \bar{x}] + \sum_{s=1}^{A_i(t)} \rho^{A_i(t) - s}\, \varepsilon_{\tau_i(t)+s}$. Conditional on $I_i(t)$, the first two terms are known and so have zero conditional variance. Hence, $Var(X_t \mid I_i(t)) = Var\left(\sum_{s=1}^{A_i(t)} \rho^{A_i(t) - s}\, \varepsilon_{\tau_i(t)+s}\right)$.

Since the innovations are mutually independent with variance $\sigma^2$, we can re-index the sum to obtain the familiar geometric-series form:

$$Var(X_t \mid I_i(t)) = \sigma^2 \sum_{s=0}^{A_i(t)-1} \rho^{2s} = \begin{cases} \sigma^2 \dfrac{1 - \rho^{2A_i(t)}}{1 - \rho^2}, & |\rho| < 1, \\ A_i(t)\sigma^2, & \rho = 1. \end{cases}$$

Predictive uncertainty depends on three factors: environmental volatility ($\sigma^2$), state persistence ($\rho$), and information age ($A_i(t)$). For a given $\sigma^2$ and $\rho \neq 0$, predictive uncertainty increases with $A_i(t)$ as unobserved innovations accumulate. The rate of accumulation depends on $|\rho|$: when $|\rho|$ is large past innovations remain persistent and uncertainty accumulates more strongly with age. In contrast when $|\rho|$ is small, the effects of past innovations decay rapidly and predictive uncertainty approaches its limiting value more quickly. For $|\rho| < 1$, predictive uncertainty is bounded above by the stationary variance $S^2 \equiv \sigma^2/(1 - \rho^2)$ and converges to it as $A_i(t) \to \infty$.[3] The predictive variance at age $a$ can thus be written as $v(a) = S^2(1 - \rho^{2a})$, with $v(a) \to S^2$ as $a \to \infty$.

By contrast, under the random walk process, predictive uncertainty about the current payoff-relevant state increases linearly with $A_i(t)$, without bound. Each additional period without an information arrival contributes exactly one innovation variance ($\sigma^2$) to the agent's uncertainty about the current payoff-relevant state.

*5.2 Information Divergence*

Under the linear Gaussian AR(1) specification, agent $i$'s predictive distribution $\mu_i(t)$ is Gaussian. Since the innovations have mean zero, its conditional mean is $\mathbb{E}[X_t \mid I_i(t)] = \bar{x} + \rho^{A_i(t)}[X_{\tau_i(t)} - \bar{x}]$, with conditional variance as derived above. Therefore, $\mu_i(\mathrm{t}) = N(\mathbb{E}[X_t \mid I_i(t)], Var(X_t \mid I_i(t)))$.

**Proposition 3 (Age Separation under Gaussian State Dynamics)**

For the stationary AR(1) process with $0 < |\rho| < 1$ and $\sigma^2 > 0$, different information ages generate distinct predictive distributions. The same result holds for the random-walk benchmark.

**Proof.** Under the AR(1) process, an agent whose most recent observation is $x$ and whose information age is $a$ has predictive distribution: $Q^a(\cdot| \ x) = N(\bar{x} + \rho^a(x - \bar{x}), v(a))$.

For $0 < |\ \rho\ | < 1$ and $\sigma^2 > 0$, $v(a)$ is strictly increasing in $a$. Hence if $a \neq a'$, then $v(a) \neq v(a')$. So the corresponding Gaussian predictive distributions must differ. Therefore, the stationary AR(1) transition kernel is age-separating.

For the random walk benchmark, the $a$-step predictive distribution is $Q^a(\cdot| \ x) = N(x, a\sigma^2)$. If $a \neq a'$, then, since $\sigma^2 > 0$, $a\sigma^2 \neq a'\sigma^2$. The predictive variances differ and so the corresponding Gaussian predictive distributions are distinct. Hence, the random-walk transition kernel is also age-separating. □

[3] In the special case $\rho = 0$, predictive uncertainty equals $\sigma^2$ for any $A_i(t) \geq 1$, so the upper bound is attained immediately rather than approached asymptotically.

Proposition 3 therefore verifies the age-separation condition underlying Theorem 1 for both the stationary Gaussian AR(1) process and the random walk benchmark. In particular, for agents $i \neq j$, $A_i(t) \neq A_j(t) \Longrightarrow \mu_i(t) \neq \mu_j(t)$. Under the random walk, agent $i$'s predictive distribution takes the especially simple form: $\mu_i(t) = N(X_{t-A_i(t)}, A_i(t)\sigma^2)$.

**Numerical Illustration:** To illustrate Theorem 2, we simulate $N$ = 6 ex ante identical agents observing a common AR(1) payoff-relevant state over $T$ = 300 periods, with identical initial information, forecasting rules, and information-arrival probabilities. We set $\rho = 0.85$, normalize the unconditional mean to $\bar{x} = 0$, and innovation standard deviation to $\sigma = 1$. Each agent independently receives fresh information with common probability $p = 0.15$ per period. This corresponds to an expected interval of approximately $1/p \approx 6.7$ periods between observations. Information arrivals are independent across agents and over time. Figure 2 illustrates the results.

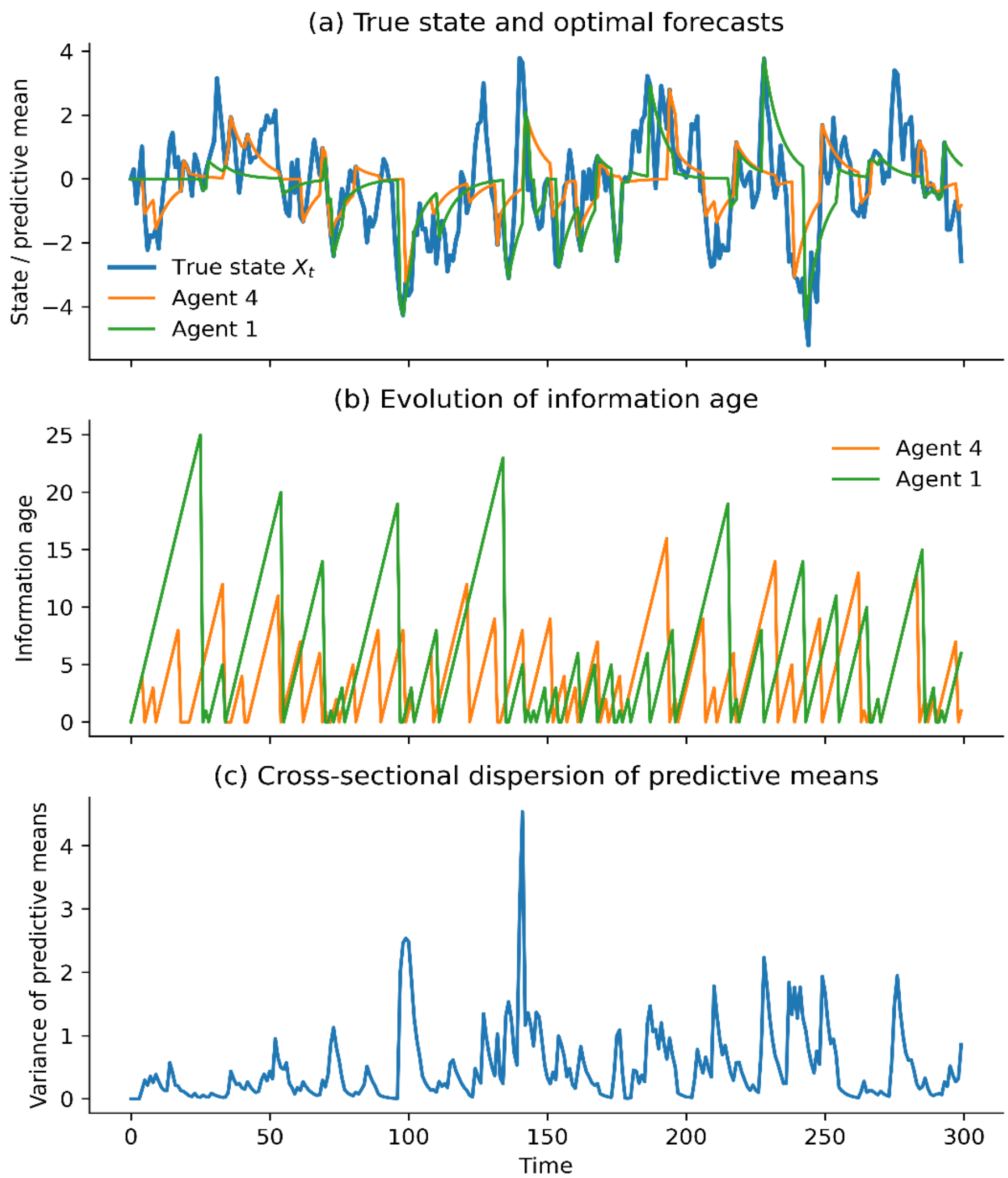

**Figure 2. Numerical illustration under an AR(1) process**

**Panel (a)** plots the true state and the predictive means of two representative agents. Although both use the same forecasting rule, asynchronous arrivals generate different forecasts: the agent with more recent information tracks the state more closely, while the forecast based on older information reverts toward the long-run mean.

**Panel (b)** plots the corresponding information-age processes. Age resets to zero when fresh information arrives and otherwise increases by one, producing the characteristic sawtooth pattern. Identical arrival probabilities therefore still generate heterogeneous realized information ages.

**Panel (c)** reports the cross-sectional dispersion of predictive means across all agents. Dispersion repeatedly rises and falls as information arrives asynchronously, illustrating the recurrent information divergence established by Theorem 2. Alternative values of $\rho$, $p$, and $N$ affect the frequency and magnitude of observed divergence but leave the underlying mechanism unchanged.

*5.3 Common Staleness, Forecast Disagreement, and Disclosure*

Asynchronous observations generate forecast disagreement, but its magnitude depends on both the information-age gap and the common staleness of the observations. These same dimensions determine the incentive to certify information age when counterparties value forecast accuracy.

Consider the stationary Gaussian AR(1) process with $0 < |\rho| < 1$. By Assumption 1, observation times (whether exogenous arrivals or chosen updates) are independent of the state process. An agent whose latest observation is $a$ periods old has conditional mean forecast $\hat{X}_t(a) = \overline{x} + \rho^a(X_{t-a} - \overline{x})$ and predictive variance $v(a)$ from Section 5.1.

Consider two information ages, $m$ and $m + g$, where $m \geq 0$ measures common staleness and $g \geq 1$ the age gap. Define expected squared forecast disagreement, conditional on these ages, as $D(m, g) = \mathbb{E}\big[(\hat{X}_t(m) - \hat{X}_t(m+g))^2\big]$.

To examine disclosure incentives, suppose an agent privately knows whether its information is fresh or stale. Before contracting, the counterparty observes neither the agent's forecast nor its underlying observation. The counterparty values the agent's service according to expected quadratic forecasting loss: $V(a) = V_0 - v(a)$, where $V_0$ denotes the value of a perfectly accurate current-state estimate. A fresh agent can verifiably certify its information age at cost $k > 0$; a stale agent cannot obtain this certificate. Certification reveals age but not the observed state. Payments equal the counterparty's posterior expected valuation.

**Proposition 4 (Common Staleness and Information-Age Disclosure)**

Under the stationary Gaussian AR(1) benchmark: $D(m,g) = v(m+g) - v(m) = S^2\rho^{2m}(1 - \rho^{2g})$. Under quadratic forecast loss, this quantity also equals the fresh type's valuation advantage. A separating equilibrium in which the fresh type certifies and the stale type remains silent exists if and only if $k \leq D(m,\ g)$.

For a fixed age gap $g$, increasing both information ages by $h \geq 1$ periods strictly increases both types' predictive uncertainty but reduces disagreement and the maximum certification cost consistent with separation: $D(m+h,g) = \rho^{2h}D(m,g) < D(m,g)$. Consequently, for any fixed $g \geq 1$ and any $k > 0$, sufficiently large common staleness $m$ rules out a separating equilibrium.

**Proof.** Stationarity implies $\mathrm{Var}\left(\hat{X}_t(a)\right) = S^2\rho^{2a}$, and $\mathrm{Cov}\left(\hat{X}_t(m),\ \hat{X}_t(m+g)\right) = S^2\rho^{2(m+g)}$. The forecasts have the same unconditional mean. Expanding their expected squared difference therefore gives $D(m,g) = S^2\rho^{2m} - S^2\rho^{2(m+g)} = v(m+g) - v(m)$.

In a separating equilibrium, silence identifies the stale type. The fresh type receives $V(m) - k$ from certification and $V(m+g)$ from silence, so certification is optimal when $k \leq D(m,g)$. The stale type cannot imitate certification. Finally, $v(a+h) - v(a) = S^2\rho^{2a}(1-\rho^{2h}) > 0$, while $D(m+h,g) = \rho^{2h}D(m,g)$ converges to zero as $h$ increases. □

Under the random-walk benchmark, $v(a) = \sigma^2 a$, so both expected squared disagreement and the certification threshold equal $D(m,g) = v(m+g) - v(m) = \sigma^2 g$, independently of $m$. Common staleness then has no effect on either quantity. The decline in disagreement and in the certification incentive with common staleness in Proposition 4 is therefore a consequence of mean reversion rather than of information age alone. The equality between the two quantities, by contrast, does not depend on mean reversion, as the following remark shows.

*Remark (A General Identity).* The identity $D(m,g) = v(m+g) - v(m)$ does not depend on the AR(1) structure. For any Markov state process with finite second moments, let $\hat{X}_F = \mathbb{E}[X_t \mid X_{t-m}]$ and $\hat{X}_S = \mathbb{E}\left[X_t \mid X_{t-m-g}\right]$ denote the fresh and stale forecasts. By the Markov property, $\hat{X}_F = \mathbb{E}\left[X_t \mid X_{t-m}, X_{t-m-g}\right]$, so by the law of iterated expectations, $\hat{X}_S = \mathbb{E}\left[\hat{X}_F \mid X_{t-m-g}\right]$. The fresh forecast error $X_t - \hat{X}_F$ is therefore orthogonal to $\hat{X}_F - \hat{X}_S$, and

$$\mathbb{E}\left[\left(X_t - \hat{X}_S\right)^2\right] = \mathbb{E}\left[\left(X_t - \hat{X}_F\right)^2\right] + \mathbb{E}\left[\left(\hat{X}_F - \hat{X}_S\right)^2\right].$$

Expected squared disagreement therefore always equals the stale type's excess mean squared forecast error. What is specific to mean reversion is the comparative static in $m$: decreasing one-period increments imply decreasing disagreement for every fixed positive age gap. Under the stationary AR(1) process, these increments equal $v(m+1) - v(m) = S^2\rho^{2m}(1-\rho^2)$, which decline geometrically in $m$. For a general Markov process, $v(a)$ is read as the unconditional mean squared forecast error at age $a$. Under the AR(1) process this coincides with the predictive variance, which does not depend on the last observed state.

**Corollary 2 (Stale Consensus)**

Under the stationary AR(1) process, fix an information-age gap $g > 0$. As common staleness $m$ increases, expected squared forecast disagreement vanishes while both agents' predictive uncertainties converge to the stationary variance:

$$D(m, g) = S^2\rho^{2m}(1-\rho^{2g}) \to 0, \text{ while } v(m) \to S^2 \text{ and } v(m+g) \to S^2 \text{ as } m \to \infty.$$

Therefore forecast disagreement can be arbitrarily small while uncertainty is arbitrarily close to its stationary maximum $S^2$.

**Proof.** From Proposition 4, the disagreement $D(m, g) = S^2\rho^{2m}(1 - \rho^{2g})$. Since $0 < |\rho| < 1$, $\rho^{2m} \to 0$ as $m \to \infty$, so $D(m, g) \to 0$ for any fixed $g$. By Section 5.1, both $v(m)$ and $v(m + g)$ converge to the stationary variance $S^2$. □

***Interpretation.*** The corollary formalizes a distinction between agreement and information quality. Two agents may issue nearly identical forecasts either because both possess very current information or because both rely on sufficiently stale observations and therefore forecast primarily from the stationary mean. In the latter case, consensus reflects the loss of state-specific information rather than informational precision. This is a *stale-consensus effect*: forecast disagreement falls while individual uncertainty rises. Observed agreement is therefore not, by itself, sufficient evidence that agents are well informed. Moreover, because Proposition 4 shows that the incentive to certify relative freshness also declines with common staleness, markets may exhibit simultaneously high uncertainty, low disagreement, and little information-age disclosure.

The model also yields a testable implication: holding observed forecast dispersion fixed, forecast errors should be larger when the information underlying those forecasts is older on average. This prediction could be examined using professional forecast vintages, analyst update dates, or timestamps from automated forecasting and AI retrieval systems.

**Numerical example.** Let $S^2 = 1$ and $\rho = 0.8$, which imply innovation variance $\sigma^2 = 0.36$. Also let $g = 2$ and $k = 0.20$. Table 1 reports predictive uncertainty $v(a)$ and expected squared disagreement $D(m, g)$ at two levels of common staleness, $m$. At $m = 1$, $k < D(1, 2)$, so the fresh type certifies and separation is sustainable. At $m = 4$, both agents are less informed, yet disagreement falls below $k$ and separation breaks down.

**Table 1. Common Staleness, Predictive Uncertainty, and Information-Age Disclosure**

| Common staleness ($m$) | $v$ (fresh) | $v$ (stale) | $D(m, 2)$ | Separation at $k = 0.2$ |
|---|---|---|---|---|
| 1 | 0.3600 | 0.7379 | 0.3779 | Yes |
| 4 | 0.8322 | 0.9313 | 0.0991 | No |

The example illustrates Proposition 4 and Corollary 2: increasing common staleness, $m$, raises predictive uncertainty while reducing both disagreement and the incentive to certify freshness.

*5.4 Information-Age Certification and Partial Unraveling*

Proposition 4 treats information age as a binary type—an agent is either fresh or stale. In many applications, however, information age is naturally ordered over more than two values: a financial statement may be one, two, or several quarters old; a credit report may have been updated days or months ago; and a forecasting service may rely on data of varying vintages. This section extends the disclosure analysis to a setting with many possible information ages.

Let information age $A \in \{0, 1, 2, \ldots\}$ be privately known. The counterparty values an agent with age $a$ at $V(a) = V_0 - v(a)$, where $v(a)$ is the forecast-error variance at age $a$, and pays its posterior expected valuation. Consider a threshold disclosure strategy: for some integer $m^* \geq 0$, agents with $A \leq m^*$ certify their exact information age, while agents with $A > m^*$ remain silent. Let $\bar{V}(m^*) = \mathbb{E}[V(A) \mid A > m^*]$ denote the counterparty's expected valuation of an uncertified agent under such a strategy.

**Proposition 5 (Partial Unraveling of Information Age)**

Suppose $V(a)$ is strictly decreasing in $a$, $\Pr(A = a) > 0$ for every integer $a \geq 0$, certification truthfully reveals exact information age at cost $k > 0$, and payments equal the counterparty's posterior expected valuation. Then,

(i) A threshold equilibrium with cutoff $m^*$ exists if and only if $V(m^* + 1) - \bar{V}(m^*) \leq k \leq V(m^*) - \bar{V}(m^*)$.

(ii) Under the stationary AR(1) process with $0 < |\rho| < 1$, $V(a) - V(a + 1) = S^2\rho^{2a}(1 - \rho^2) \to 0$ as $a \to \infty$. Moreover, every equilibrium threshold satisfies $k < S^2\rho^{2m*}$, and hence $m^* < \ln(S^2/k)/\ln(\rho^{-2})$. Thus, for every $k > 0$, all sufficiently stale information ages pool in every threshold equilibrium. If $k \geq S^2$, no threshold equilibrium with positive certification exists. Full disclosure is never an equilibrium when $k > 0$ and the support of information age is unbounded.

**Proof.** (i) Consider first a type with age $a \leq m^*$. Certification truthfully reveals age $a$ and yields $V(a) - k$. If the type instead remains silent, it joins the pool $A > m^*$ and receives $\bar{V}(m^*)$. Because $V(a)$ is decreasing in $a$, the certifying type with the weakest incentive to certify is $a = m^*$. Hence all certifying types prefer certification if and only if $V(m^*) - k \geq \bar{V}(m^*)$, or equivalently $k \leq V(m^*) - \bar{V}(m^*)$.

Now consider a silent type with $a > m^*$. If it deviates and certifies, it receives $V(a) - k$. The silent type with the strongest incentive to deviate is the freshest silent type, $a = m^* + 1$. Hence all silent types prefer silence if and only if $\bar{V}(m^*) \geq V(m^* + 1) - k$, or equivalently $k \geq V(m^* + 1) - \bar{V}(m^*)$.

Because $\Pr(A > m^*) > 0$, silence occurs on the equilibrium path and $\bar{V}(m^*)$ is pinned down by Bayes' rule. Combining the two incentive conditions gives the stated bounds.

(ii) Under the stationary AR(1) process, $V(a) = V_0 - S^2 + S^2\rho^{2a}$. Therefore $V(a) - V(a + 1) = S^2\rho^{2a}(1 - \rho^2) \to 0$. For every finite $m^*$, the silent pool contains only ages strictly greater than $m^*$, so $\bar{V}(m^*) > V_0 - S^2$. The upper incentive condition in part (i) therefore implies $k \leq V(m^*) - \bar{V}(m^*) < S^2\rho^{2m*}$.

Solving the strict upper bound for $m^*$ gives $m^* < \ln(S^2/k)/\ln(\rho^{-2})$. Hence any positive certification cost imposes a finite upper bound on the ages that can certify in a threshold equilibrium. If $k \geq S^2$, the upper incentive condition cannot hold for any $m^*$.

Finally, suppose every type certifies. Silence is then off the equilibrium path, but any posterior belief over feasible information ages assigns a silent agent a valuation no lower than $V_0 - S^2$. Since $V(a) \to V_0 - S^2$, for every $k > 0$ there exist sufficiently large ages $a$ such that $V(a) - k < V_0 - S^2$. Such a type strictly prefers to deviate to silence. Full disclosure therefore cannot be an equilibrium. □

*Interpretation.* The mechanism is a temporal analogue of standard unraveling. Fresher types have an incentive to separate from the silent pool, but mean reversion limits complete unraveling by compressing valuation differences as information becomes stale. Proposition 5(ii) makes this precise: for every $k > 0$, there is an age beyond which certification cannot occur in any threshold equilibrium. Costly certification therefore produces partial rather than complete unraveling—not because old observations were inaccurate when recorded, but because the incremental value of distinguishing sufficiently stale vintages becomes too small relative to the certification cost.

Proposition 5 extends the costly-disclosure logic of Jovanovic (1982) and Verrecchia (1983) to an ordered set of information ages. The distinctive feature is that the state dynamics themselves compress valuation differences as information becomes stale. The proposition characterizes threshold equilibria but does not establish uniqueness; full pooling may coexist with threshold equilibria for some certification costs.

*Geometric Information Ages.* Under Bernoulli observation with per-period probability $u$—whether interpreted as the exogenous arrival probability of Section 4.2 or as the updating intensity chosen in Section 4.3—the stationary information-age distribution is $\Pr(A = a) = u(1 - u)^a$, $a = 0, 1, 2, \ldots$ . Define $\theta(u) \equiv u / [1 - (1 - u)\rho^2]$.

By the memoryless property of the geometric distribution, conditional on silence under cutoff $m$, $\Pr(A = m + 1 + j \mid A > m) = u(1 - u)^j$, $j = 0, 1, 2, \ldots$ . Hence $\mathbb{E}[\rho^{2A} \mid A > m] = \rho^{2(m+1)} \theta(u)$, and the pooled valuation of silent types is $\bar{V}(m) = V_0 - S^2 + S^2\rho^{2(m+1)}\theta(u)$.

**Corollary 3 (Threshold Equilibria under Geometric Information Ages)**

Under the stationary AR(1) process with geometric information ages generated by updating probability $0 < u < 1$, define $k_L(m;\, u) = S^2\rho^{2(m+1)}[1 - \theta(u)]$ and $k_H(m;\, u) = S^2\rho^{2m}[1 - \rho^2\theta(u)]$. A threshold $m$ is an equilibrium cutoff if and only if $k_L(m;\, u) \le k \le k_H(m;\, u)$.

(i) A threshold equilibrium exists for every $0 < k \le k_H(0;\, u)$. If $k > k_H(0;\, u)$, no threshold equilibrium with positive certification exists. Full pooling is an equilibrium if and only if $k \ge S^2[1 - \theta(u)]$. Consequently, full pooling and a positive-certification threshold equilibrium coexist whenever $S^2[1 - \theta(u)] \le k \le k_H(0;\, u)$.

(ii) For any $k$ for which a threshold equilibrium exists, the set of equilibrium thresholds is a nonempty set of consecutive integers. Define $m_L(k;\, u) = \min\{m \ge 0 : k_L(m;\, u) \le k\}$ and $m_H(k;\, u) = \max\{m \ge 0 : k \le k_H(m;\, u)\}$. Then the equilibrium thresholds are exactly $\{m_L(k;\, u), \ldots, m_H(k;\, u)\}$. As $k \downarrow 0$, both $m_L(k;\, u)$ and $m_H(k;\, u)$ diverge to infinity.

(iii) Both $k_L(m;\, u)$ and $k_H(m;\, u)$ are strictly decreasing in $u$. Consequently, for fixed $k > 0$, $m_L(k;\, u)$ and $m_H(k;\, u)$ are weakly decreasing in $u$ over the range where $k \le k_H(0;\, u)$, so that positive-certification threshold equilibria exist. Within this range, more frequent information arrival raises the valuation of the silent pool and weakly lowers both the smallest and largest equilibrium cutoffs.

**Proof.** (i) Fix $u \in (0,1)$. By the geometric memoryless property, the pooled valuation under cutoff $m$ is $\overline{V}(m) = V_0 - S^2 + S^2\rho^{2(m+1)}\theta(u)$. Substituting this expression into the two incentive conditions in Proposition 5(i) gives $k_L(m;u) = S^2\rho^{2(m+1)}[1-\theta(u)]$ and $k_H(m;u) = S^2\rho^{2m}[1-\rho^2\theta(u)]$. Because $0 < \theta(u) < 1$, both bounds are strictly positive and decline geometrically with $m$ at rate $\rho^2$. Moreover, $k_H(m;u) - k_L(m;u) = S^2\rho^{2m}(1-\rho^2) > 0$, and $k_H(m+1;u) - k_L(m;u) = S^2\rho^{2(m+1)}\theta(u)(1-\rho^2) > 0$. Thus the admissible intervals overlap consecutively. Since both endpoints converge to zero as $m \to \infty$, their union is

$$\bigcup_{m=0}^{\infty}[k_L(m;u), k_H(m;u)] = (0, k_H(0;u)].$$

Consequently, a threshold equilibrium exists for every $0 < k \le k_H(0;u)$. If $k > k_H(0;u)$, no threshold satisfies the upper incentive condition, so no threshold equilibrium with positive certification exists. Under full pooling, silence yields the payment $\mathbb{E}[V(A)] = V_0 - S^2 + S^2\theta(u)$.

Because $V(a)$ is strictly decreasing, the freshest type, $a = 0$, has the strongest incentive to deviate by certifying. Full pooling is therefore an equilibrium if and only if $V(0) - k \le \mathbb{E}[V(A)]$. Since $V(0) = V_0$, this condition is equivalent to $k \ge S^2[1-\theta(u)]$. Furthermore, $k_H(0;u) - S^2[1-\theta(u)] = S^2\theta(u)(1-\rho^2) > 0$. Hence full pooling and a positive-certification threshold equilibrium coexist whenever $S^2[1-\theta(u)] \le k \le k_H(0;u)$.

(ii) Fix $k$ such that $0 < k \le k_H(0;u)$. Because both bounds are strictly decreasing in $m$, the lower incentive condition holds exactly when $m \ge m_L(k;u)$, whereas the upper incentive condition holds exactly when $m \le m_H(k;u)$. The existence result in part (i) ensures that these conditions hold simultaneously for at least one integer $m$. Thus $m_L(k;u) \le m_H(k;u)$, and the equilibrium cutoffs form the nonempty consecutive set $\{m_L(k;u), \dots, m_H(k;u)\}$.

To establish divergence as $k \downarrow 0$, fix any integer $M \ge 0$. Since $k_L(M;u) > 0$, sufficiently small $k$ satisfies $k < k_L(M;u)$. Because $k_L(m;u)$ decreases with $m$, no cutoff $m \le M$ then satisfies the lower incentive condition. Therefore $m_L(k;u) > M$. As $M$ was arbitrary, $m_L(k;u) \to \infty$ as $k \downarrow 0$. Since $m_H(k;u) \ge m_L(k;u)$, the largest equilibrium cutoff also diverges to infinity.

(iii) For $0 < u < 1$, $\theta'(u) = (1-\rho^2)/\,[1-(1-u)\rho^2]^2 > 0$. It follows that, for every fixed $m$, $\partial k_L(m;u)/\,\partial u = -S^2\rho^{2(m+1)}\theta'(u) < 0$ and $\partial k_H(m;u)/\,\partial u = -S^2\rho^{2(m+1)}\theta'(u) < 0$.

For fixed $k > 0$, increasing $u$ therefore weakly enlarges the set $\{m \ge 0 : k_L(m;u) \le k\}$ and weakly shrinks the set $\{m \ge 0 : k \le k_H(m;u)\}$. Consequently, both $m_L(k;u)$ and $m_H(k;u)$ are weakly decreasing in $u$ over the range where $k \le k_H(0;u)$, so that positive-certification threshold equilibria exist.

**Numerical Example.** Continue with $S^2 = 1$ and $\rho = 0.8$, normalize $V_0 = 1$ so that $V(a) = 0.8^{2a}$, and set the certification cost to $k = 0.10$. By Proposition 5(ii), no age above $\ln(1/0.10)/\ln(1/0.64) \approx 5.16$ can certify in any threshold equilibrium, whatever the distribution of information ages. Table 2 reports the equilibrium cutoffs under geometric ages for three updating intensities; $u = 0.25$ corresponds to one update per year on average in quarterly data.

**Table 2. Equilibrium disclosure under geometric information ages**
($S^2 = 1$, $\rho = 0.8$, $k = 0.10$)

| Updating intensity $u$ | $\theta(u)$ | Equilibrium cutoffs $m$* | Share certifying, Pr($A \leq m$*) |
|---|---|---|---|
| 0.10 | 0.2358 | 4 | 0.41 |
| 0.25 | 0.4808 | 3, 4 | 0.68, 0.76 |
| 0.50 | 0.7353 | 2, 3 | 0.88, 0.94 |

Three features stand out. First, unraveling is partial: every equilibrium cutoff lies below the bound of 5.16, and all older ages pool. Second, equilibria need not be unique: with $u = 0.25$, both $m^* = 3$ and $m^* = 4$ are equilibrium cutoffs. Third, more frequent updating lowers the cutoff, as Corollary 3(iii) predicts, from 4 at $u = 0.10$ to 2 or 3 at $u = 0.50$. Yet the share of agents who certify rises, from 41 percent to roughly 90 percent, because more agents hold fresh information. Frequent updating therefore narrows the range of ages worth certifying while increasing how often certification occurs. Figure 3 illustrates the results.

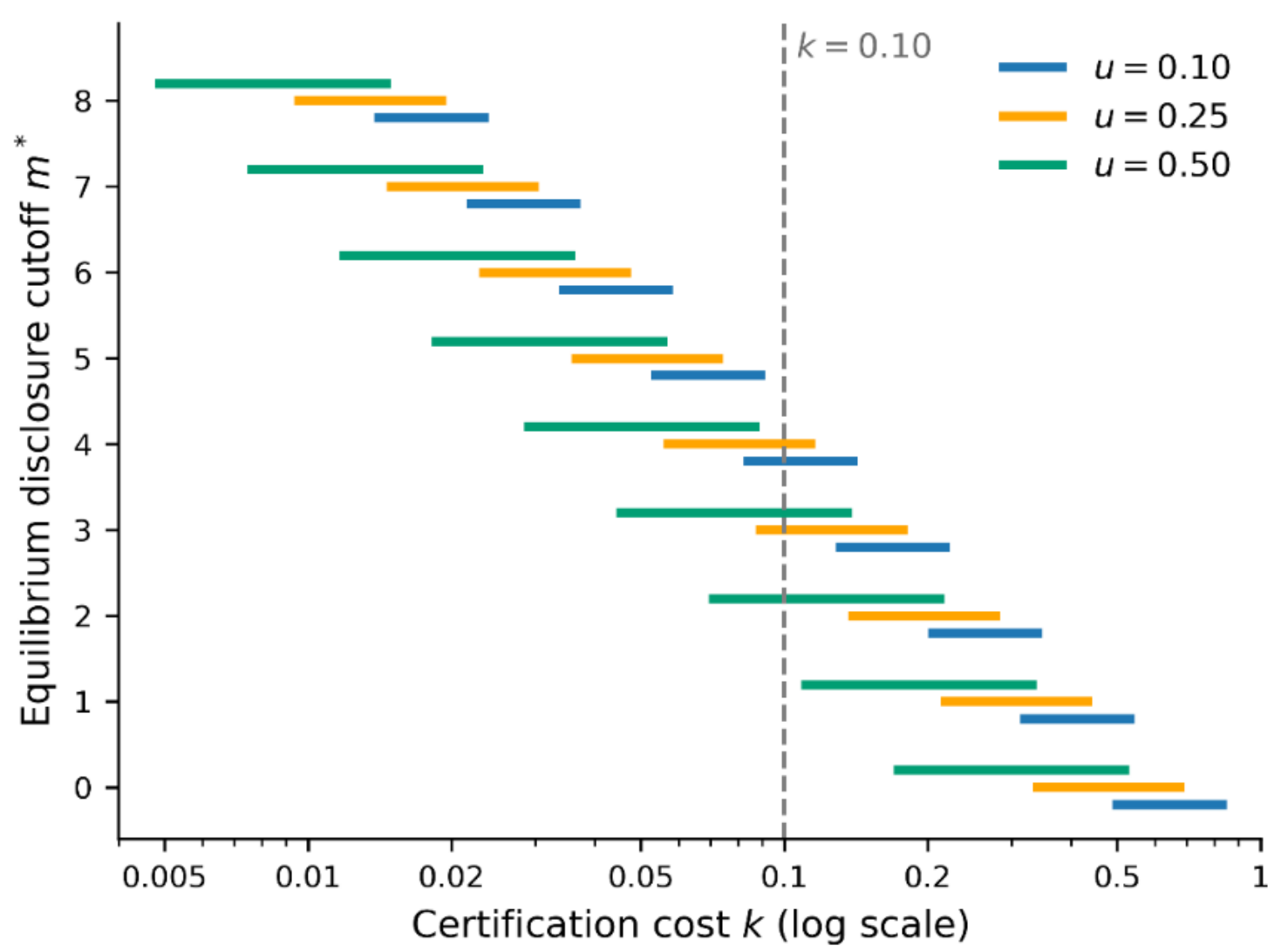

**Figure 3. Certification-cost intervals supporting threshold equilibria**

In Figure 3, each horizontal segment represents $[k_L(m;\, u),\, k_H(m;\, u)]$, with $S^2 = 1$ and $\rho = 0.8$. The dashed line marks $k = 0.10$. Vertical offsets distinguish updating intensities at the same integer cutoff. For each $u$, the range of costs supporting cutoff $m$ shifts toward lower costs as $m$ rises, and adjacent ranges overlap, so every cost up to $k_H(0; u)$ supports at least one cutoff and some costs support several. A higher $u$ shifts these ranges toward lower costs, weakly lowering the smallest and largest equilibrium cutoffs wherever positive-certification equilibria exist.

## 6. Applications

### *6.1 Credit, Insurance, and Information Certification*

Credit and insurance markets provide natural applications because borrower creditworthiness and policyholder risk evolve over time while lenders, insurers, and intermediaries often rely on records generated at different dates. Credit reports, income information, claims histories, medical records, driving records, and inspections may therefore give otherwise similar decision makers information of different ages about the same underlying risk. Even when these observations were accurate when obtained, differences in information age can generate different predictive assessments.

Mortgage underwriting illustrates the practical relevance of information vintage. Fannie Mae and Freddie Mac impose explicit age limits on several forms of underwriting documentation and require older information to be updated or reverified before closing (Fannie Mae, 2025; Freddie Mac, 2025). These rules do not implement the equilibrium mechanism derived in Section 5.4, but they show that institutions treat the age of otherwise valid information as economically relevant. Open banking provides a related application: consent-based access to recent borrower records may reduce differences in information age across lenders, while verifiable observation timestamps could lower the cost of establishing freshness.

The common-staleness result implies that lenders or insurers relying on increasingly old records may form more similar assessments even as uncertainty about current risk rises. Apparent agreement across institutions may therefore reflect common staleness rather than precise current information. Recency can also have disclosure value: borrowers, policyholders, intermediaries, or information providers may benefit from certifying that favorable information is relatively recent, while the value of distinguishing among sufficiently old records may fall below verification costs.

### *6.2 Financial Markets and Forecasting*

Financial markets provide another natural application because investors and analysts routinely rely on information generated at different times. Earnings announcements, regulatory filings, analyst research, proprietary data, and macroeconomic releases may be accurate when produced but differ in age when used. Two otherwise similar analysts can therefore form different forecasts simply because they rely on different information vintages rather than different models or signal precision.

The common-staleness result implies that low forecast dispersion need not indicate high information quality. When analysts rely on relatively recent but differently dated information, age differences may generate substantial disagreement. As their information becomes commonly stale, however, forecasts may converge toward similar long-run expectations even while uncertainty about current fundamentals remains high.

This issue is recognized in practice by consensus-estimate providers, which screen forecasts partly by age. LSEG's StarMine SmartEstimate methodology weights forecasts by analyst accuracy and recency and excludes stale estimates, while I/B/E/S removes certain forecasts before constructing consensus estimates (Kaplan, Martin, and Xie, 2021). Information intermediaries may likewise have incentives to disclose update dates, data vintages, and research timestamps when recency materially affects predictive value.

### *6.3 Communication Networks and Multi-Agent AI*

In satellite, sensor, and remote-monitoring systems, transmission constraints, intermittent connectivity, congestion, or failed observations can leave different receivers with updates of different ages. Age of Information has consequently become an important engineering measure of data freshness, including in low-Earth-orbit satellite communication systems (Kaul, Yates, and Gruteser, 2012; Yates et al., 2021; Zhang et al., 2025).

Our framework adds an informational implication: when several decision makers rely on asynchronous observations of the same evolving state, heterogeneous information ages can generate heterogeneous predictive distributions, even when all observations are exact and all agents use the same model. By the common-staleness result, receivers may also agree because all are operating from old information, not because the underlying state is known precisely.

The same logic applies to multi-agent AI systems, in which specialized agents retrieve documents, query databases, or observe external environments at different times. An orchestrator that monitors only agreement across agents may mistake stale consensus for reliable consensus. Tracking source-observation and retrieval timestamps and data provenance alongside agent beliefs can distinguish agreement based on current evidence from agreement generated by common staleness.

## 7. Conclusion

This paper develops a dynamic theory of the endogenous emergence of asymmetric information. Its central mechanism is simple: otherwise identical rational agents can become differently informed when they observe the same evolving payoff-relevant state at different times. Because information age changes the predictive distribution of the current state, asynchronous observation generates heterogeneous effective information states and, under an age-separating transition kernel, divergent predictive distributions. Asymmetric information can therefore arise without exogenously endowed private information, heterogeneous signals, or differential information-acquisition technologies.

These informational differences need not be temporary. Under independent, nondegenerate stochastic information arrival, information divergence recurs infinitely often almost surely even when agents face identical arrival processes. When updating is costly, heterogeneous information ages can also arise endogenously as agents trade off the cost of obtaining fresh information against the expected loss from acting on stale information.

The linear-Gaussian results show that the consequences of age-based asymmetry depend not only on the age gap but also on common staleness. Under mean reversion, agents relying on increasingly stale observations become more uncertain individually while their forecasts become more similar. Low disagreement can therefore reflect stale consensus rather than precise information. The same mechanism shapes disclosure incentives: relatively fresh information may be worth certifying, while sufficiently stale vintages pool because the value of distinguishing nearby ages eventually falls below certification costs.

These implications are relevant wherever decisions rely on information generated at different dates, including credit and insurance underwriting, financial forecasting, and multi-agent AI systems (Section 6). The framework also bears on information-sharing policy: open banking and standardized observation timestamps could reduce age-based asymmetries and certification costs, although their welfare effects depend on updating costs, privacy, and market responses. The stale-consensus result likewise cautions against treating agreement across institutions as evidence of accurate assessments.

Several extensions follow naturally. Section 4.3 models endogenous updating through a chosen updating intensity; a next step would allow updating decisions to depend dynamically on current information age and other states, yielding a Bellman formulation for optimal information refresh. Such age-dependent rules could resynchronize identical agents, so asymmetry would persist only if the updating technology remained stochastic or agents differed. The framework could also be extended to market design, principal–agent problems, and environments in which observation timing depends on the underlying state, so that the absence of an update itself becomes informative.

More broadly, the paper shifts attention from the consequences of asymmetric information to one mechanism through which it can arise. When the underlying state evolves, when an agent observes the state becomes part of what the agent knows about it.

## References

Akerlof, G. A. (1970): "The Market for 'Lemons': Quality Uncertainty and the Market Mechanism," *Quarterly Journal of Economics*, 84, 488–500.

Andrade, P., and H. Le Bihan (2013): "Inattentive Professional Forecasters," *Journal of Monetary Economics*, 60(8), 967–982.

Ball, I. (2023): "Dynamic Information Provision: Rewarding the Past and Guiding the Future," *Econometrica*, 91, 1363–1391.

Barron, J. M., and J. Ni (2008): "Endogenous Asymmetric Information and International Equity Home Bias: The Effects of Portfolio Size and Information Costs," *Journal of International Money and Finance*, 27(4), 617–635.

Bomberger, W. A. (1996): "Disagreement as a Measure of Uncertainty." *Journal of Money, Credit and Banking*, 28(3), 381–392.

Cai, Z. (2019): "Dynamic Information Acquisition and Time-Varying Uncertainty," *Journal of Economic Theory*, 184, 104947.

Calvo, G. A. (1983): "Staggered Prices in a Utility-Maximizing Framework," *Journal of Monetary Economics*, 12(3), 383–398.

Coibion, O., and Y. Gorodnichenko (2012): "What Can Survey Forecasts Tell Us about Information Rigidities?" *Journal of Political Economy*, 120(1), 116–159.

Coibion, O., and Y. Gorodnichenko (2015): "Information Rigidity and the Expectations Formation Process: A Simple Framework and New Facts," *American Economic Review*, 105(8), 2644–2678.

Daley, B., and B. Green (2012): "Waiting for News in the Market for Lemons," *Econometrica*, 80(4), 1433–1504.

Dye, R. A. (1985): "Disclosure of Nonproprietary Information." *Journal of Accounting Research*, 23(1), 123–145.

Ely, J. C. (2017): "Beeps," *American Economic Review*, 107, 31–53.

Fannie Mae (2025): "*Selling Guide*, Section B1-1-03, Allowable Age of Credit Documents and Federal Income Tax Returns." Washington, DC: Federal National Mortgage Association. Available at selling-guide.fanniemae.com.

Fershtman, C., and A. Pakes (2012): "Dynamic Games with Asymmetric Information: A Framework for Empirical Work," *Quarterly Journal of Economics*, 127, 1611–1661.

Freddie Mac (2025): *Single-Family Seller/Servicer Guide*, Section 5102.4, "Age of Documentation." McLean, VA: Federal Home Loan Mortgage Corporation. Available at guide.freddiemac.com.

Gonzalez, P. (2004): “Investment and Screening under Asymmetric Endogenous Information,” *RAND Journal of Economics*, 35(3), 502–519.

Grossman, S. J. (1981): “The Informational Role of Warranties and Private Disclosure about Product Quality,” *Journal of Law and Economics*, 24(3), 461–483.

Grossman, S. J., and O. D. Hart (1983): “An Analysis of the Principal–Agent Problem,” *Econometrica*, 51, 7–45.

Grossman, S. J., and J. E. Stiglitz (1980): “On the Impossibility of Informationally Efficient Markets,” *American Economic Review*, 70, 393–408.

Hébert, B., and J. La'O (2023): “Information Acquisition, Efficiency, and Nonfundamental Volatility,” *Journal of Political Economy*, 131(10), 2666–2723.

Hellwig, C., and L. Veldkamp (2009): “Knowing What Others Know: Coordination Motives in Information Acquisition,” *Review of Economic Studies*, 76, 223–251.

Holmström, B. (1979): “Moral Hazard and Observability,” *Bell Journal of Economics*, 10, 74–91.

Jovanovic, B. (1982): “Truthful Disclosure of Information.” *Bell Journal of Economics*, 13(1), 36–44.

Kamenica, E., and M. Gentzkow (2011): “Bayesian Persuasion,” *American Economic Review*, 101, 2590–2615.

Kaplan, Z., X. Martin, and Y. Xie (2021): “Truncating Optimism,” *Journal of Accounting Research*, 59(5), 1827–1884.

Kaul, S., R. D. Yates, and M. Gruteser (2012): “Real-Time Status: How Often Should One Update?” *Proceedings of IEEE INFOCOM*, 2731–2735.

Lahiri, K., and X. Sheng (2010): “Measuring Forecast Uncertainty by Disagreement: The Missing Link,” *Journal of Applied Econometrics*, 25(4), 514–538.

Mankiw, N. G., and R. Reis (2002): “Sticky Information versus Sticky Prices: A Proposal to Replace the New Keynesian Phillips Curve,” *Quarterly Journal of Economics*, 117, 1295–1328.

Mankiw, N. G., R. Reis, and J. Wolfers (2003): “Disagreement about Inflation Expectations,” *NBER Macroeconomics Annual*, 18, 209–248.

Matějka, F., and A. McKay (2015): “Rational Inattention to Discrete Choices: A New Foundation for the Multinomial Logit Model,” *American Economic Review*, 105(1), 272–298.

Milgrom, P. R. (1981): “Good News and Bad News: Representation Theorems and Applications,” *Bell Journal of Economics*, 12(2), 380–391.

Milgrom, P., and J. Roberts (1986): “Price and Advertising Signals of Product Quality,” *Journal of Political Economy*, 94, 796–821.

Mitra, A., J. A. Richards, S. Bagchi, and S. Sundaram (2022): “Distributed State Estimation over Time-Varying Graphs: Exploiting the Age of Information,” *IEEE Transactions on Automatic Control*, 67(12), 6349–6365.

Myatt, D. P., and C. Wallace (2012): “Endogenous Information Acquisition in Coordination Games,” *Review of Economic Studies*, 79, 340–374.

Ouyang, Y., H. Tavafoghi, and D. Teneketzis (2025): “An Approach to Stochastic Dynamic Games with Asymmetric Information and Hidden Actions,” *Dynamic Games and Applications*, 15(1), 182–215.

Reis, R. (2006): “Inattentive Producers,” *Review of Economic Studies*, 73, 793–821.

Renault, J., E. Solan, and N. Vieille (2017): “Optimal Dynamic Information Provision,” *Games and Economic Behavior*, 104, 329–349.

Rothschild, M., and J. E. Stiglitz (1976): “Equilibrium in Competitive Insurance Markets: An Essay on the Economics of Imperfect Information,” *Quarterly Journal of Economics*, 90, 629–649.

Sims, C. A. (2003): “Implications of Rational Inattention,” *Journal of Monetary Economics*, 50, 665–690.

Smolin, A. (2021): “Dynamic Evaluation Design,” *American Economic Journal: Microeconomics*, 13, 300–331.

Spence, M. (1973): “Job Market Signaling,” *Quarterly Journal of Economics*, 87, 355–374.

Steiner, J., C. Stewart, and F. Matějka (2017): “Rational Inattention Dynamics: Inertia and Delay in Decision-Making,” *Econometrica*, 85, 521–553.

Subedi, G., and R. K. Shrestha (2026): “Information Latency: Theory and Economic Consequences,” SSRN, http://dx.doi.org/10.2139/ssrn.7497238 or ArXiv:2609.31951.

Veldkamp, L. (2006): “Information Markets and the Comovement of Asset Prices,” *Review of Economic Studies*, 73, 823–845.

Verrecchia, R. E. (1983): “Discretionary Disclosure,” *Journal of Accounting and Economics*, 5, 179–194.

Yates, R. D., Y. Sun, D. R. Brown III, S. K. Kaul, E. Modiano, and S. Ulukus (2021): “Age of Information: An Introduction and Survey,” *IEEE Journal on Selected Areas in Communications*, 39, 1183–1210.

Zarnowitz, V., and L. A. Lambros (1987): “Consensus and Uncertainty in Economic Prediction,” *Journal of Political Economy*, 95(3), 591–621.

Zhang, D., Y. Peng, Y. Ke, J. Song, S. Wang, G. Pan, and J. An (2025): “Age of Information Variation of LEO Satellite-Terrestrial Uplink Transmissions,” *IEEE Transactions on Vehicular Technology*, 74(6), 9645–9655.

Zhong, W. (2022): “Optimal Dynamic Information Acquisition,” *Econometrica*, 90(4), 1537–1582.